\documentclass[%
 aip,
 jmp,%
 amsmath,amssymb,
 reprint,%
]{revtex4-1}

\draft 

\usepackage{graphicx}
\usepackage{dcolumn}
\usepackage{bm}

\begin{document}





\title{Enhanced emission and light control with tapered plasmonic nanoantennas}








\author{Ivan S. Maksymov}
\email{mis124@physics.anu.edu.au}
\author{Arthur R. Davoyan}
\email{ard124@physics.anu.edu.au}
\author{Yuri S. Kivshar}
\affiliation{Nonlinear Physics Centre and Centre for Ultrahigh Bandwidth Devices for Optical Systems (CUDOS), Research School of Physics and Engineering, \\ Australian National University, Canberra ACT 0200, Australia}





\date{\today}

\begin{abstract}
We introduce a design of Yagi-Uda plasmonic nanoantennas for enhancing the antenna gain and achieving control over the angular emission of light. We demonstrate that tapering of antenna elements allows to decrease spacing between the antenna elements tenfold also enhancing its emission directivity. We find the optimal tapering angle that provides the maximum directivity enhancement and the minimum end-fire beamwidth.
\end{abstract}

\pacs{78.67.Bf, 42.70.Qs, 42.82.Et, 71.45.Gm}

\maketitle 

Plasmonic nanoantennas are used to couple free-propagating radiation to subwavelength confined regions~\cite{bha09,nov11}, enhance and quantize the emission of single quantum emitters such as single molecules or single quantum dots~\cite{huc11,mak10}, and offer flexible control over the directionality of nanofocused light~\cite{she11}.

Recently, rapid progress has been made in the realization of directional control over radiation from single quantum emitters by means of plasmonic Yagi-Uda antennas composed of appropriately arranged metal nanoparticles \cite{cur10,tam08,kos10, hof07,fem09}. High directivity of such plasmonic antennas originates from the uniqueness of their construction inspired from radio frequency (RF) technology. A classical RF Yagi-Uda antenna \cite{bal05, antenna_book} employs mutual coupling between standing-wave current elements to produce a traveling-wave unidirectional pattern. It uses parasitic elements around an active feed element for reflector and directors to produce an end-fire beam.

In order to fully explore the potential of the communication of energy to, from, and between single quantum emitters, plasmonic Yagi-Uda antennas should have higher antenna gain and offer control mechanisms over the directivity. These aims can be achieved by increasing the length of the antenna along the end-fire beam direction by incrementing the number of antenna directors.

While the control over the directionality does not present significant difficulties at RF frequencies, at optical frequencies absorption losses impose severe restrictions on the length of the antenna. Specifically, the optical response of noble metals such as gold or silver is given by their complex dielectric constants characterized by a negative real part and a positive imaginary part. The latter describes the absorption of radiation energy in the material \cite{boz09}. Since the electric field of localized  surface plasmon modes penetrates into the elements of the antenna, metal absorption losses, which are a function of the electric field intensity and the imaginary part of the dielectric function of the constituent material, increase as increases the number of directors, and the effect from the antenna lengthening comes to naught.

The restriction on the length and consequently the directivity of the plasmonic antenna imposed by the inevitable metal losses in the additional directors can be overcome by slowly tapering the length of the directors. The improvement of the directivity of RF Yagi-Uda antennas by slowly varying the length of directors was suggested by Sengupta \cite{sen60} in 1959. But his idea has been abandoned in the modern RF antenna technique because the suggested tapering can be applied to long Yagi-Uda antennas only and results in heavy constructions. At optical frequencies, tapering of the plasmonic waveguides \cite{ner97,bab00,sto04,gra05,dav10a} and metamaterials \cite{roc09,muh10} was suggested as an effective way for nanofocusing of plasmons. The effect of the taper shape and the optimal taper angle on nanofocusing has been discussed actively in the literature (see e.g Ref. ~\onlinecite{dav10}). However the application of tapered plasmonic waveguides as directional antennas is challenging because the coupling of single emitters to plasmonic modes of the untapered waveguide end is not efficient \cite{cha06,*cha07}.

In this Letter, we suggest an approach for directional control of plasmonic nanoantennas by employing tapered waveguides, and suggest a design principle of plasmonic Yagi-Uda antennas for the substantial enhancement of the antenna gain and control of its directivity. We demonstrate that, in contrast to classical design principles, the spacing between the elements of plasmonic nanoantennas can be decreased tenfold and the directivity can be enhanced by slowly tapering the elements length. We obtain numerically the optimal tapering angle that provides the maximum directivity enhancement and the minimum end-fire beamwidth.

For the purpose of demonstration, we consider a two-dimensional model of a 5-element and a 42-element silver plasmonic Yagi-Uda antennas embedded in air and designed according to the effective wavelength rescaling approach \cite{nov07} relying on a linearly scaled effective wavelength introduced to account for antenna geometry and metal properties at optical frequencies. This rescaling offers additional degrees of freedom in designing antennas with shorter element lengths and inter-element spacings than those used in classical RF antennas.

The schematics of the considered 5-element antenna is shown in Fig. $1$(a). The antenna consists of a resonant reflector, the feed of and equally spaced directors. The length of the feed is $L_{f}=\lambda_{0}/4$, where $\lambda_{0}=1550$ nm is the emitted wavelength. The reflector element of length $L_{r}=1.25L_{f}$ is inductively detuned to resonate at wavelengths longer than the resonate wavelength of the feed. The directors of length $L_{f}=0.9L_{f}$ are capacitively detuned to resonate at wavelengths shorter than the resonate wavelength of the feed. The width of all elements is $a=50$ nm and the spacing between all element is $w=30$ nm. According to Refs. ~\onlinecite{cur10, tam08}, the quantum emitter (pointlike, with a vacuum wavelength $\lambda_{0}$) having an in-plane \textit{x}-polarized dipole is placed near one of the ends of the feed element. This layout ensures a strong near-field coupling to the feed mode because the position of the emitter coincides with a position of high electric mode density \cite{cha06,*cha07}.

\begin{figure}
\includegraphics[width=6cm]{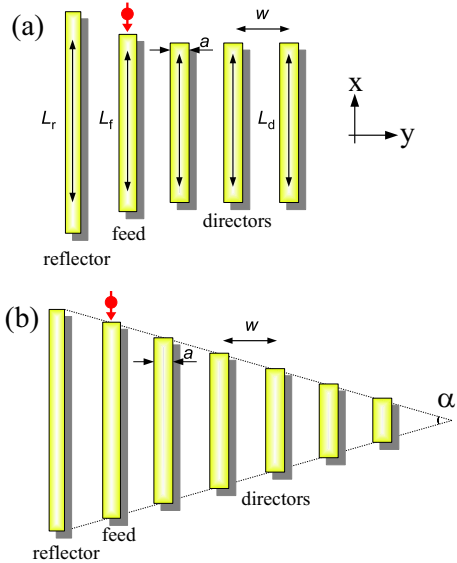}
\caption{\label{fig:epsart} (a) Schematic view of a 5-element plasmonic Yagi-Uda antenna consisting of a reflector of length $L_{r}$, a feed of length $L_{f}$ and three directors of length $L_{d}$. The spacings between the reflector and the feed and the directors are $w$. The red arrow denotes the position and the polarization of the emitter. (b) Schematic of the tapered plasmonic Yagi-Uda antenna with length of the director, feed and directors chosen according to the tapering angle $\alpha$. All the elements of the tapered antenna are equidistant with spacing $w$.}
\end{figure}

The angular dependence of the emission of both antennas is studied first. The finite-element method (FEM) \cite{com11} is used to simulate the electromagnetic fields inside and between the elements of the antenna as well as the fields propagated away from the antenna. Our numerical model takes into consideration the electric fields lying in the $XY$-plane (Fig. $1$). The dielectric function of silver used in the simulations is based on a Drude model fit for the published optical constants of silver \cite{pal85}. Simulations were performed for different Yagi-Uda architectures by altering the number and dimensions of directors and the spacings between the element of the antenna.

According to Ref.~\onlinecite{bal05}, the performance of the investigated antennas is discussed in terms of the angular directivity defined as

\begin{eqnarray}
D(\theta)=\frac{2{\pi}P(\theta)}{P_{loss}+P_{rad}}
\label{eq:one}.
\end{eqnarray}

Here $P(\theta)$ denotes the angular radiated power, $P_{rad}$ denotes the Poynting vector flux through a closed contour in the far-field region and $P_{loss}$ is the absorption losses proportional to $\epsilon''|\textbf{E}|^2$, where $\epsilon''$ is the imaginary part of the dielectric function of silver at $1550$ nm. All electromagnetic quantities involved into Eq.~(\ref{eq:one}) are obtained with FEM simulations.

\begin{figure}
\includegraphics[width=8.5cm]{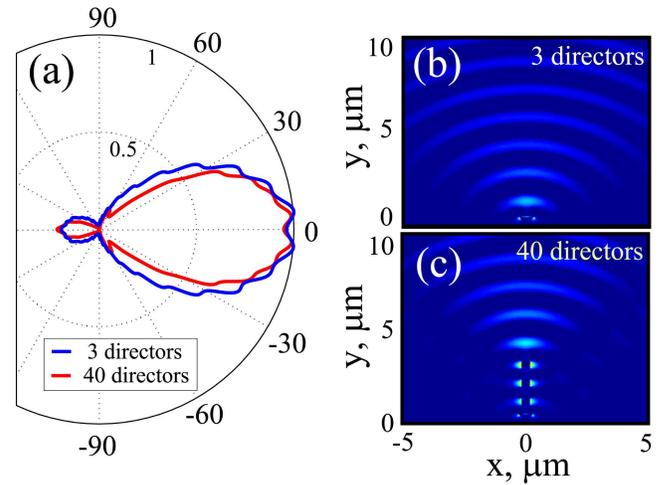}
\caption{\label{fig:epsart} (a) Simulated angular emission patterns of the 5-element (blue line) and 42-element (red line) Yagi-Uda antennas. (b, c) Real part of $E_{x}$ electric fields of the 5-element and 42-element Yagi-Uda antennas. Red and blue colors correspond to the minimum and maximum field strengths, respectively.}
\end{figure}

Figure~$2$(a) shows the emission pattern of the 5-element (blue line) and 42-element (red line) antennas. It demonstrates that the 5-element antenna with the spacings between the elements of just $30$ nm offers a typical directivity as compared to a similar antenna designed according classical design principles \cite{tam08}. Simulations for the 42-element antenna with the same element dimensions and inter-element spacings but with an increased number of directors demonstrate that the increase in the total length of the antenna does not change the directivity and just slightly decreases the width of the main lobe. Indeed, as shown in Fig. $2$ (a), the maximum emission directivity [defined as $D_{max}=max(D(\theta))$] for the 42-element antenna equals to that of the 5-element antenna. The analysis of the near-field profiles of the generated end-fire beams [Figs. $2$ (b) and (c)] convincingly confirms a poor performance of the elongated antenna. This result contradicts the predictions of the classical antenna theory \cite{antenna_book} because the benefit from the augmentation of the number of directors is brought to naught by absorption losses in the metallic elements.

Now we turn our attention to the tapered antenna [Fig. $2$(b)]. The length of the feed and the length of the reflector of the tapered antenna, the width of all elements and the spacings between them equal to the corresponding dimensions of the antennas without the tapering [Fig. $1$ (a)]. The length of the directors is slowly tapered as a function of the tapering angle $\alpha$ according to the formula $L(\alpha)=L_{f}-2(W+a)tg(\alpha)$, where $L_{f}$ is the feed length, $a$ is the width of the directors and $W=Nw$ is the spatial coordinate of the center of the $N^{th}$ director. 

\begin{figure}
\includegraphics[width=8.5cm]{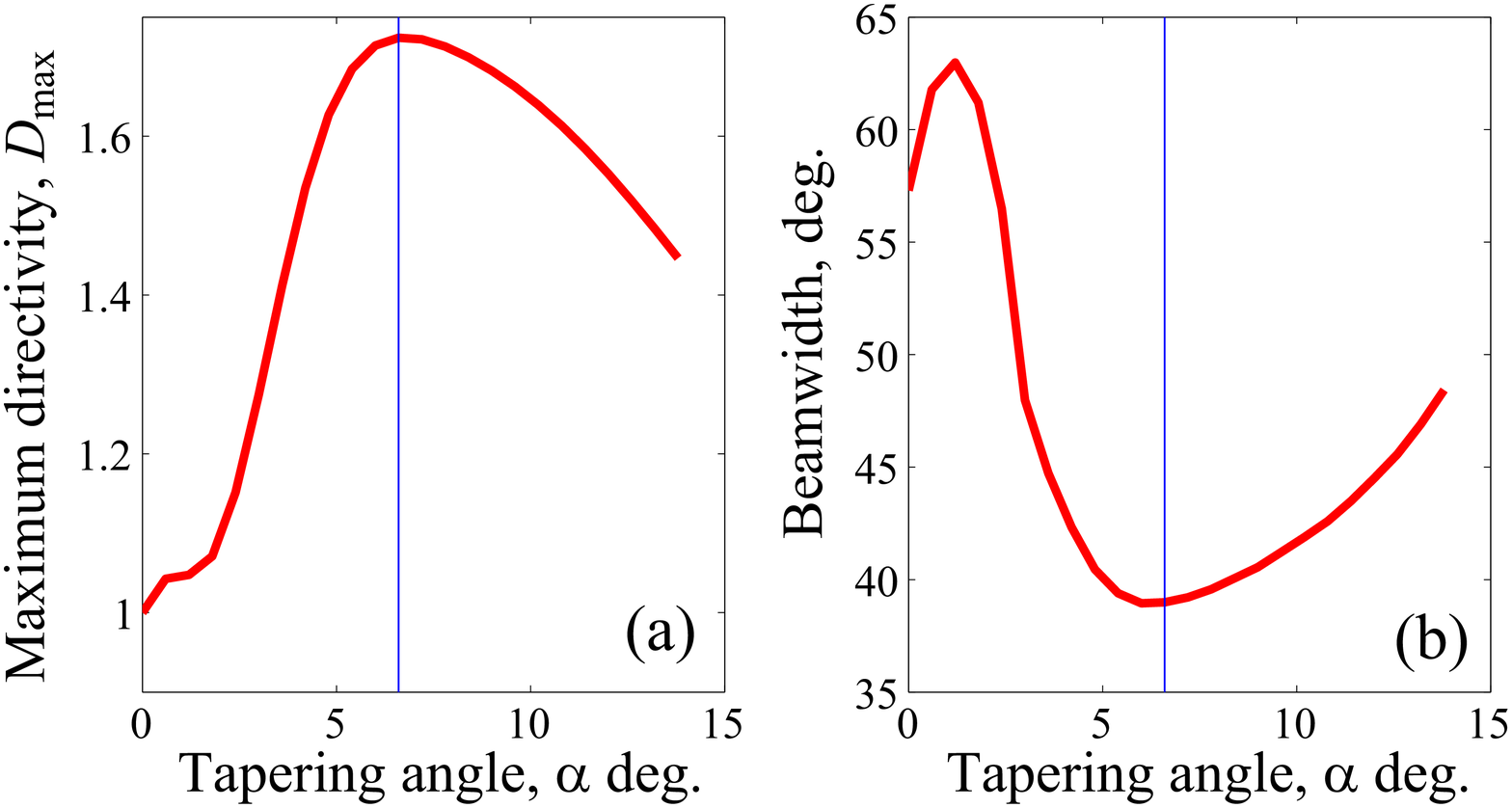}
\caption{\label{fig:epsart} (a) Maximum emission directivity $D_{max}$ of the tapered 42-element Yagi-Uda antenna as a function of the tapering angle $\alpha$. (b) Beamwidth of the tapered 42-element Yagi-Uda antenna as a function of the tapering angle $\alpha$. The straight blue lines highlight the optimal tapering angle $\alpha_{opt}=6.6^{o}$.}
\end{figure}

Fig. $3$ (a) plots the maximum emission directivity $D_{max}$ normalized to the maximum directivity of the 42-element antenna without tapering [blue line in Fig. $2$(a)] as a function of the tapering angle $\alpha$. The beamwidth of the antenna, defined as the angular separation between two identical points on opposite side of the pattern half-maximum \cite{bal05}, is plotted in Fig. $3$ (b) also as a function of $\alpha$. A pronounced maximum of the directivity [Fig. $3$ (a)] and a minimum of the beamwidth [Fig. $3$ (b)] are observed at $\alpha_{opt}=6.6^{o}$, which is hereafter referred to as the optimal tapering angle.

In contrast to the antenna without tapering ($\alpha_{opt}=0$), the optimally tapered antenna offers a more than one-and-a-half superior directivity accompanied by a $\sim20^{o}$ decrease in the beamwidth, which is a considerable advantage that originates from the effect of the optimal tapering \cite{dav10} accompanied by a decrease in absorption losses due to the reduction of the metallic elements. Despite a further decrease in absorption offered by the antenna with $\alpha=12.2^{o}$, its directivity deteriorates as compared to the antenna with the optimal tapering angle (however the directivity of the antenna with $\alpha=12.2^{o}$ is still superior than that of the antenna without tapering).

In order to demonstrate the impact of the tapering on the directivity, we repeat the calculations of the angular emission patterns and the near-field distributions for three different 42-element Yagi-Uda antennas without tapering ($\alpha=0$), with the optimal tapering angle ($\alpha_{opt}=6.6^{o}$) and with a higher than optimal tapering angle ($\alpha=12.2^{o}$), respectively. The calculated emission patterns are shown in Fig. $4$ (a). According to the result in Fig. $3$, in Fig. $4$ (a) we observe a $\sim1.7$ increase in the maximum directivity (red line) and a substantial narrowing of the end-fire beam. A broadened beam pattern (green line) confirms the prediction of Fig. $3$ at tapering angles $\alpha > \alpha_{opt}$.

The calculated near-field profiles [Figs $4$ (b,c)] reveal that at the optimal tapering angle the field propagates to the tapered end of the antenna and a narrow end-fire beam is formed. In contrast, at tapering angles $\alpha > \alpha_{opt}$ the field is not confined to the antenna directors and jumps off before it reaches the end of the antenna.

These results convincingly confirm that the tapering with the optimal angle allows enhancing the directivity of the 42-element plasmonic Yagi-Uda antenna. We remind that this result is achieved with a $30$ nm spacing between the antenna elements, which equals to just one tenth of the spacing of state-of-the-art plasmonic Yagi-Uda antennas \cite{cur10, kos10}. We notice that the use of optimization techniques \cite{jon97} will facilitate the design of tapered antennas embedded in different background materials as well as further improve the performance characteristics.

\begin{figure}
\includegraphics[width=8.5cm]{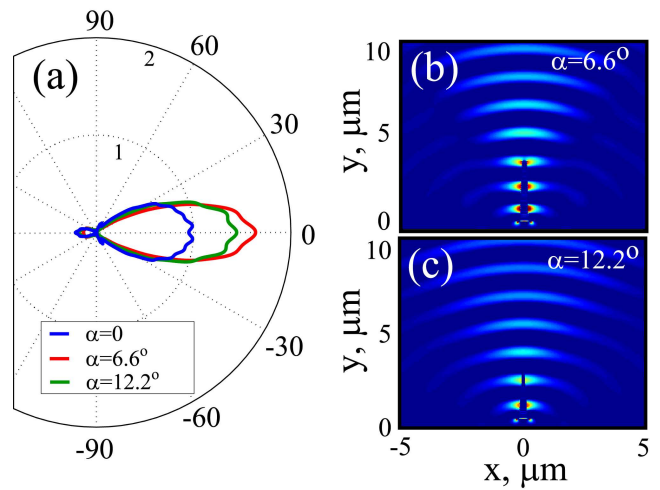}
\caption{\label{fig:epsart} (a) Simulated angular emission patterns of 42-element Yagi-Uda antennas without tapering (blue line, $\alpha=0$), with the optimal tapering (red line, $\alpha_{opt}=6.6^{o}$), and with a non-optimized tapering (green  line,  $\alpha=12.2^{o}$). Notice a difference in the radius-vector limits adopted in Fig. $2$(a). All patterns are normalized to the maximum directivity of the antenna without tapering. (b,c) Real part of $E_{x}$ electric fields of the antenna with the optimal tapering ($\alpha_{opt}=6.6^{o}$) and with a non-optimized tapering ($\alpha=12.2^{o}$). The same colormap and colorscale as in Figs. $2$ (b,c) are used.}
\end{figure}

In conclusion, we have applied the concept of tapered plasmonic waveguides to plasmonic antennas and suggested the enhanced emission directivity and reduction of the total nanoantenna length. Our results suggest that the application of the antenna design principles to optical nanoantennas cannot be straightforward, and the use of plasmonic nanoparticles as active and passive antenna elements can pave ways for effective control over the subwavelength light. We expect that a very small footprint of the proposed antennas will make them especially attractive for application to horizontal on-chip plasmonic wireless signal-transmission circuits.

The authors thank Andrey Miroshnichenko and Ilya Shadrirov for valuable discussions. This work was supported by the Australian Research Council.

\providecommand{\noopsort}[1]{}\providecommand{\singleletter}[1]{#1}%

\end{document}